\documentclass[12pt]{article}

\usepackage{amsmath}
\usepackage{epsfig}


\usepackage{thophys}
\usepackage{thohacks}

\begin{document}
\thispagestyle{empty}
\begin{titlepage}
\begin{flushright}
hep-ph/9805297 \\
TTP 98--16 \\
HD--THEP 98--18 \\
\today
\end{flushright}

\vspace{0.3cm}
\boldmath
\begin{center}
\Large\bf  Evolution of the Light--Cone Distribution Function for a 
           Heavy Quark
\end{center}
\unboldmath
\vspace{0.8cm}

\begin{center}
{\large Christopher Balzereit} and {\large Thomas Mannel} \\
{\sl Institut f\"{u}r Theoretische Teilchenphysik,
     Universit\"at Karlsruhe,\\ D -- 76128 Karlsruhe, Germany} 
\vspace*{5mm} \\
{\large Wolfgang Kilian} \\
{\sl Institut f\"{u}r Theoretische Physik,
     Universit\"at Heidelberg,\\ D -- 69120 Heidelberg, Germany} 
\end{center}

\vspace{\fill}

\begin{abstract}
\noindent
We compute the one-loop anomalous dimension for the light cone 
distribution function of a heavy quark and solve the corresponding 
evolution equation analytically. Some implications of the results 
for inclusive $B$ decays are discussed.   
\end{abstract}
\end{titlepage}

\newpage

\section{Introduction}
\label{sec:introduction}

 The heavy mass expansion for inclusive decays of heavy hadrons
has become a generally accepted tool. Although there are hints  
that there could be problems in the calculation of lifetimes
(the $\Lambda_b$--lifetime) within this framework, 
the $1/m_Q$ expansion seems to be an 
appropriate method to access semileptonic
processes\cite{FaLuSav1,FaLuSav2,FaLu}. 

However, if one is aiming at a complete description of the 
spectra (i.e. energy of the charged lepton $E_\ell$ or the 
hadronic invariant mass $\sqrt{Q_H^2}$) in general problems 
are encountered in the endpoint region, where $Q_H^2$ becomes 
small. Typically this invariant mass is of the order $m_B$, the 
mass of the decaying hadron. This is true in almost all phase 
space; however, getting closer to the endpoint it has become 
popular to distinguish two regions. The first region is the one
with very small 
$Q_H^2 \sim \Lambda_{QCD}^2$ and also very small hadronic 
energy $v \cdot Q_H \sim \Lambda_{QCD}$ (here $v$ is 
the velocity of the decaying hadron); here the proper description is 
a summation over the very few exclusive channels contributing in 
this region. 

However, in the second region where the hadronic energy is still large 
$v \cdot Q_H \sim m_B$ but the invariant mass $Q_H^2$ becomes small, 
namely $Q_H^2 \sim \Lambda_{QCD} m_B$, there is still the 
possibility to obtain a proper QCD description of the decay. 
In this case it has been shown \cite{Neu,ManNeu,Bigietal}
that one may resum the leading twist terms into a universal light-cone 
distribution function (or shape function) for the heavy quark.
This function is entirely non-perturbative and may be written 
formally as a forward matrix element of a non-local operator.
This nonlocal operator corresponds to the Fourier transform 
of the Wilson line \cite{KorSter,KorGroz}.

On the other hand, there are computable perturbative corrections 
which exhibit logarithmic singularities close to the endpoint 
region. To first order in $\alpha_s$ the leading contribution to 
the spectrum close to the endpoint is of 
the form $\alpha_s \ln (Q^2) / Q^2$, where $Q^2$ is now the partonic 
analogue of the hadronic invariant mass of the final state. Upon 
integration this yields the usual Sudakov logarithms. 

Several methods have been proposed to combine both perturbative and 
non-perturbative contributions. At least for the doubly logarithmic 
terms this should be possible since they are process independent and 
hence could be assigned to the universal shape function. In addition, 
close to the endpoint, a resummation of the double logarithms becomes 
necessary, since these terms become large. Such a resummation may be 
achieved by setting up an evolution equation for the shape function 
as it has been proposed in \cite{NeuGroz,KorGroz}. 

In the present note we use such an evolution equation and solve it 
analytically by putting in the one loop result for the anomalous 
dimensions.
In the next section we express the hadronic invariant mass spectra
in terms of the shape function, thereby defining a limit in which 
the leading twist dominates. In section 3 we consider the renormalization 
of the shape function and set up an evolution equation and derive the 
anomalous dimension in section 4. Finally we solve the evolution equation 
analytically and discuss the implications for the evolution of the 
moments of the shape function.


\section{Large energy limit of hadronic invariant mass spectra}
\label{sec:spectra}

In inclusive decays such as $B\rightarrow X_s + \gamma$
and $B \rightarrow X_u+ l + \nu_l$
the effective hamiltonian takes the general form
\begin{equation}
\mathcal H_{eff} = j_{\mu}H^{\mu},
\end{equation}
where $H^{\mu}$ is a hadronic current and $j^{\mu}$ is either leptonic or
photonic.  

We shall in the following consider the invariant mass spectrum of the 
final state hadrons. We denote with $Q_H^2$ the square of this invariant 
mass. The spectrum may be written as 
\begin{equation}
\frac{d\Gamma}{dQ_H^2} = \frac{1}{2m_B}
\int d\Phi_\ell \sum_X \int d\Phi_X 
(2\pi)^4\delta^4(P - q - P_X) \delta(Q_H^2 - P_X^2)
L_{\mu\nu}H^{\mu\nu},
\end{equation}
where phase space integration over the leptonic and hadronic total momentum
$q$, $P_X$ is denoted by $d\Phi_{\ell, X}$, respectively.

The leptonic (or photonic) and hadronic tensors are defined as
\begin{eqnarray}
L_{\mu\nu} &=& \langle 0 | j_{\mu} |l \rangle \langle l | j_{\nu}^{\dagger}
| 0 \rangle \\
H_{\mu\nu} &=& \langle B(P) | H_{\mu} | X \rangle \langle X |
H_{\nu}^{\dagger} | B(P) \rangle \,.
\end{eqnarray}
The non-hadronic  part can be calculated perturbatively
and is treated in the following as a known function of $q$.
The object of our interest is the hadronic part,
\begin{equation}
  Q_{\mu\nu} = \sum_X\int d\Phi_X (2\pi)^4\delta^4(P-q-P_X)H_{\mu\nu},
\end{equation}
which in general cannot be determined perturbatively.
Proceeding along standard lines we have 
\begin{equation}
Q_{\mu\nu} = \int d^4x \, e^{iqx}
\langle B(P) | H_{\mu}^{\dagger}(0)H_{\nu}(x)| B(P) \rangle
\end{equation}
The hadronic current is usually a bilinear function of two quark operators
one of which is a heavy $b$ quark. Using the fact that the heavy mass $m_b$
sets a large scale compared to $\Lambda_{QCD}$ we may set up an operator 
product expansion in the usual way. We write
\begin{equation}
H_{\mu} = \bar q(x) \Gamma_{\mu} b(x) 
= e^{-im_b vx} \bar q(x) \Gamma_{\mu} b_v(x),
\end{equation}
where $q(x)$ represents  a massless quark and 
the large part of the $b$--quark momentum has been explicitely removed 
from the field by a phase redefinition. This leaves us with 
\begin{equation}
Q_{\mu\nu} = \int d^4x \, e^{-ix(m_b v - q)}
\langle B(P) | \bar b_v(0) \Gamma^{\dagger}_{\mu} q(0)\,
\bar q(x)\Gamma_{\nu} b_v(x) | B(P) \rangle,
\end{equation}

In the following we are interested in the endpoint region, which 
is defined by a specific kinematical limit denoted large--energy limit.
We shall define this 
limit first in terms of the partonic total momentum of the final state
\begin{equation}
  Q^\mu = m_b v^\mu - q^\mu
\end{equation}
corresponding to the partonic invariant mass $Q^2$ and the partonic energy
$v\cdot Q$.
In the endpoint region the total energy $v\cdot Q$ of the final state scales
with the heavy quark mass $m_b$ while the invariant mass $Q^2$ is of $\mathcal
O(\Lambda_{QCD}m_{b})$ such that the light--cone component $k_+$ of the total
final state momentum remains finite of $\mathcal O(\Lambda_{QCD})$:
\begin{eqnarray} 
m_b &\rightarrow& \infty \nonumber \\
v\cdot Q &\rightarrow& \infty \nonumber \\
\frac{2v\cdot Q}{m_b} &=& \text{const.} \label{limit}  \\
k_+ &=& -\frac{Q^2}{2v \cdot Q} =  \text{const.} \nonumber  \\
\frac{Q^2}{m_b^2}&\to& 0 
\nonumber
\end{eqnarray} 
We shall relate the partonic variables to the hadronic ones at the 
end of the section.  

It has been shown~\cite{ManNeu} that in the limit (\ref{limit}) the leading
contribution may be obtained by contracting the light quark
\begin{eqnarray}\label{Wilson}
Q_{\mu\nu} &=& \int d^4x \int \frac{d^4k}{(2\pi)^4}\Theta(k_0)\delta(k^2)
e^{-ix(Q -k)}\nonumber\\
&&
\langle B(P) | \bar b_v (0) \Gamma^{\dagger}_{\mu}\fmslash{k} \Gamma_{\nu}
\mathcal{P} \exp\left[-i\int_0^x dx\cdot A(x)\right]
b_v (x) | B(P) \rangle .
\end{eqnarray}
This expression may be Taylor-expanded around $x=0$ and the expansion resummed
to yield formally
\begin{align}
Q_{\mu\nu} =&  \int \frac{d^4k}{(2\pi)^4}\Theta(k_0)\delta(k^2)\nonumber\\
&\langle B(P) | \bar b_v(0) \Gamma_{\mu}^{\dagger}\fmslash{k}\Gamma_{\nu}
(2\pi)^4\delta^4(Q - k + iD)b_v(0) | B(P) \rangle\,.
\end{align}
In the following, $\delta$--functions as in the preceding expression are
defined by their Fourier transform, which involves a path-ordered
exponential as in~(\ref{Wilson}).

Performing the $k$--integration we get
\begin{equation}
\label{eq:Qcontr}
Q_{\mu\nu} = \langle B(P) | \bar b_v(0) \Gamma_{\mu}^{\dagger}
(\fmslash{Q} + i\fmslash{D})\Gamma_{\nu}\delta((Q+ iD)^2)\,
b_v(0) | B(P) \rangle.
\end{equation} 

In the endpoint region (\ref{limit}) $Q$ is almost light--like and can 
be decomposed as 
\begin{equation}
\label{eq:ldef}
Q = (v\cdot Q)n_+ - k \quad \text{with} \quad v\cdot n_+ = 1,
\quad n_+^2 =0,\quad
k = \mathcal O(\Lambda_{QCD}) \,.
\end{equation}   
Thus in (\ref{eq:Qcontr}) we can approximate ($iD_+\equiv n_+\cdot iD$)
\begin{eqnarray*}
(Q + iD)^2 &\approx& Q^2 + 2v\cdot Q iD_+ \\
\fmslash{Q} + i\fmslash{D} &\approx& \fmslash{Q}\\
b_v &\approx& h_v\,,
\end{eqnarray*}
neglecting terms of $\mathcal O(\Lambda_{QCD}/m_b)$:
\begin{equation}
\label{eq:strufu1}
\hat Q_{\mu\nu} = \langle B(v) | \bar h_v(0) \Gamma_{\mu}^{\dagger}\fmslash{Q}
\Gamma_{\nu} \delta(Q^2 + 2v\cdot Q iD_+) h_v(0) | B(v) \rangle 
\end{equation} 
For the perturbative calculation in the next section it is useful to 
rewrite $\hat Q_{\mu\nu}$ in terms of the imaginary part 
\begin{equation}
\hat Q_{\mu\nu} = \mathrm{Im}\frac{1}{\pi} \hat T_{\mu\nu}\,,
\end{equation}
where
\begin{equation}
\label{eq:corrLEEFT}
\hat T_{\mu\nu} = - \langle B(v) | \bar h_v(0) 
\Gamma_{\mu}^{\dagger}\fmslash{Q} \Gamma_{\nu} 
\frac{1}{Q^2 + 2v\cdot Q iD_+ + i\epsilon} h_v(0) | B(v) \rangle \,.
\end{equation}

Using heavy--quark symmetry to disentangle the Dirac structure of this 
expression we finally arrive at  
\begin{equation}
\label{eq:strufu2}
\hat Q_{\mu\nu} = \frac{1}{2} 
\mathrm{Tr}\{\Gamma_{\mu}^{\dagger}\fmslash{Q} \Gamma_{\nu} P_v^+ \}
\int dk_+ \delta(Q^2 + 2v \cdot Q k_+) f(k_+),
\end{equation}
where $P_v\equiv (1+\fmslash v)/2$.

Here we have introduced the shape function or light--cone distribution 
function for the heavy quark 
\begin{equation}
\label{eq:strufu}
f(k_+) = \langle B(v) | \bar h_v \delta(k_+ - iD_+) h_v | B(v) \rangle 
\, .
\end{equation} 
The light--cone structure function $f(k_{+})$ measures
the probability of finding a heavy quark with light--cone component
$k_{+}$ inside the B--meson and is a universal function. 

Up to now we have worked in terms of partonic variables, while 
experimentally only hadronic variables are of interest. Introducing the 
hadronic light--cone variable in terms of the hadronic invariant mass
squared $Q_H^2$ and the hadronic energy $v\cdot Q_H$ we have 
\begin{equation}
K_+ = -\frac{Q^2_H}{2v\cdot Q_H}  
    = -\frac{Q^2 + 2 \bar{\Lambda} (v\cdot Q)}{2(v\cdot Q) + 2 \bar{\Lambda} } 
    = k_+ - \bar{\Lambda} + {\cal O} (\bar{\Lambda}^2 / m_b) 
\end{equation}
we find that the support of $f(k_{+})$ is
the interval $- \infty < k_{+} < \bar \Lambda$ such that 
$K_+ \le 0$.  In reality, the spectrum is peaked near $k_+=0$.
The shape function may directly be measured in a 
inclusive semileptonic or radiative decay by measuring the spectrum of
the hadronic light--cone variable $K_+$.

\section{Renormalization of the light--cone structure function}
\label{sec:matching}
In the last section we have shown, that
in the large--energy limit the integrated hadronic
tensor $Q_{\mu\nu}$ can be approximated by the
quantity $\hat Q_{\mu\nu}$ defined in the large energy limit of QCD.
Up to now only non--perturbative corrections resummed in the 
structure function are included.

To really establish the approximation we have to check 
that this also works perturbatively,
since it is well known that in the kinematical region
of small invariant mass $Q^{2}$ 
the perturbative spectrum is plagued by large logarithmic corrections.
Therefore we should require that order by order in perturbation theory 
the leading IR singularities of $Q_{\mu\nu}$
arising in the large--energy limit  
should be reproduced by
$\hat Q_{\mu\nu}$. The leading logarithms are universal and hence we 
may assign them to be perturbative corrections to the structure 
function.

In order to  identify the relevant IR singularities which have to 
be reproduced by perturbative corrections to 
the structure function we shall first consider the 
hadronic invariant mass spectrum calculated in full QCD
in the large--energy limit.
The relevant terms of the one loop contribution can be generically 
written as
\begin{equation}
\frac{d\Gamma^{(pert)(1)}}{dQ^2} \approx(\frac{\alpha_{s}}{\pi})\biggl[
\biggl(\frac{\ln(Q^2/m_b^2)}{k_{+}}\biggr)_{+}f_{1}(Q^2/m_b^2)
+\biggl(\frac{1}{k_{+}}\biggr)_{+}f_{2}(Q^2/m_b^2)
\biggr]
\end{equation} 
where the real functions $f_{i}(\rho)$ are regular in the limit
$\rho \rightarrow 0$ with $k_+$ fixed. 
From this we easily identify the IR singularity 
as the $\ln(Q^2/m_b^2)$--term since $k_+$ is held fixed in the large--energy
limit. 
If the large--energy limit makes sense in perturbation theory, 
we have to require that exactly this term is reproduced 
by the $\mathcal O(\alpha_s)$--corrections to the
light--cone distribution function.

The origin of this logarithmic divergence in the spectrum  can be traced back 
to double logarithms of the correlator 
$T^{\mu\nu}$. Thus to one-loop order the double 
logarithmic terms in $T^{\mu\nu}$
and $\hat T^{\mu\nu}$ have to match in the large--energy limit.
It therefore suffices to look for the most singular Feynman integrals
contributing to the correlators at $\mathcal O(\alpha_s)$. 

\begin{figure}
\vspace{-.5cm}
 \begin{center}
  \leavevmode
  \epsfxsize=12.6cm
  \epsffile[60 630 510 730]{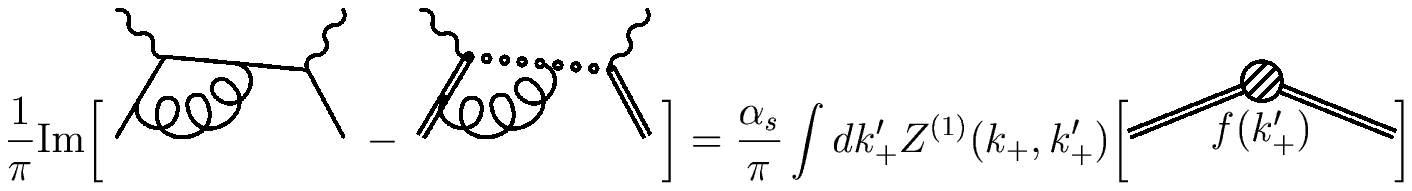}
 \end{center}
\vspace*{-1cm}
\caption{\label{fig:fig1} Renormalization of the structure function.
The plain line denotes a massive or massless QCD propagator, the double 
plain line a heavy quark propagator, the dotted
line a light--cone propagator. The shaded blob on the r.h.s. symbolizes
the insertion of the structure function as a nonlocal operator.}
\end{figure}

The one-loop corrections to the shape function consist of vertex diagrams,
self-energy contributions and a box-type diagram.  However,
in full QCD the UV divergencies of the self-energy diagrams cancel with the UV
divergency of the vertex diagram because of current conservation.  For the
shape function divergent pieces appear in the self energy of the heavy quark
which will be taken into account below.   The self-energy insertion into the
light-cone propagator does not contribute.   Thus, we can
restrict ourselves to the
vertex diagrams shown in figure~\ref{fig:fig1}.
\newpage
The part of the vertex correction in
full QCD (first term in Fig.\ref{fig:fig1}) responsible for double logarithms
reads 
\begin{eqnarray}
\label{eq:threeQCD}
V_{\rm (sing)}^{\mu\nu} &=& ig_s^2 C_F 2 
\frac{\Gamma^{\dagger\mu}\fmslash{Q}\Gamma^{\nu}}{k_+^\prime-k_++i\epsilon}\\
& &\int \frac{d^D\ell}{(2\pi)^D} 
\frac{1}{(v\cdot \ell + v\cdot k^\prime
+\frac{1}{2m}(\ell+k^\prime)^2)
\ell^2\left[(\ell+k^\prime + Q)^2+i\epsilon\right]} \nonumber \,,
\end{eqnarray}
where $k^{\prime}$ is the residual momentum of the heavy quark and 
the integration measure indicates dimensional regularization with
$D=4-\epsilon$.

On the other hand, we can directly compute the corrections to the 
shape function either by reading off the propagator from 
(\ref{eq:corrLEEFT}) or by the usual way of calculating corrections 
to Wilson lines~\cite{KorSter}. The corresponding expression is 
\begin{eqnarray}
\label{eq:threeLEEFT}
\hat V_{\rm (sing)}^{\mu\nu}
 &=& ig_s^2 C_F 2 
\frac{\Gamma^{\dagger\mu}\fmslash{Q}\Gamma^{\nu}}{k_+^\prime-k_++i\epsilon}\\
& &\int \frac{d^D\ell}{(2\pi)^D} 
\frac{1}{(v\cdot\ell+v\cdot k^\prime)
\ell^2( Q^2 + 2v\cdot Q(\ell_{+}+k_+^\prime) + i\epsilon)} \nonumber \,
\end{eqnarray}
in which the leading logarithms have to match the full QCD result 
(\ref{eq:threeQCD}).

Evaluating the corresponding integrals $I$ and $\hat I$ 
to double logarithmic 
accuracy and supressing constant terms we get
\begin{eqnarray}
\label{eq:dbllogQCD}
I &=& -\frac{i}{(4\pi)^{2}}\frac{1}{v\cdot Q}\frac{1}{2}\ln^{2}
\biggl(
\frac{k_+-k_+^\prime-i\epsilon}{\mu}
\biggr)  \\ \label{eq:dbllog}
\hat I &=& -\frac{i}{(4\pi)^{2}}(4\pi)^{\frac{\epsilon}{2}}\frac{1}{v\cdot Q}
    \Gamma(1-\epsilon)\Gamma(\epsilon)\Gamma(\frac{\epsilon}{2})
\biggl(
\frac{k_+-k_+^\prime-i\epsilon}{\mu}
\biggr)^{-\epsilon} \\ 
&=& -\frac{i}{(4\pi)^{2}}\frac{1}{v\cdot Q}
\biggl[
\frac{2}{\epsilon^{2}} - \frac{2}{\epsilon}\ln
\biggl(
\frac{k_+-k_+^\prime-i\epsilon}{\mu}
\biggr)
+ \frac{c}{\epsilon} + 
\ln^{2}
\biggl(
\frac{k_+-k_+^\prime-i\epsilon}{\mu}
\biggr)
\biggr] \nonumber\,,
\end{eqnarray}
where
\begin{equation}
c = \ln(4\pi) - \gamma_{E}  \,.
\end{equation}
Note that (\ref{eq:dbllogQCD}) is valid up to 
single logarithmic corrections.

Due to the presence of both UV and collinear divergencies, a $1/\epsilon^2$
pole appears.  However, both singularities are of UV nature in the sense that
the integral converges for $\epsilon>0$ and can therefore be removed by
renormalization, as will be explained below.

In the difference of both integrals 
\begin{eqnarray}
\label{eq:diffint}
\Delta I &=& I - \hat I \\
&=& -\frac{i}{(4\pi)^{2}}\frac{1}{v\cdot Q}
\biggl[\!
-\frac{2}{\epsilon^{2}}\! +\! \frac{2}{\epsilon}\ln
\biggl(
\frac{k_+-k_+^\prime-i\epsilon}{\mu}
\biggr)\!
- \!\frac{c}{\epsilon}\!
-\!\frac{1}{2}\ln^{2}
\biggl(
\frac{k_+-k_+^\prime-i\epsilon}{\mu}
\biggr)\!
\biggr] \nonumber
\end{eqnarray}
the double logarithms do not cancel.  

The operator to be renormalized [i.e., the operator defining the light-cone
distribution function~(\ref{eq:strufu})] is of nonlocal nature.  Hence, we can
introduce an integration over $k_+'$, which plays the role of a summation over
a continuous operator basis labeled by $k_+'$, to absorb the local as well as
the nonlocal UV
divergencies in~(\ref{eq:diffint}) into a renormalized distribution function.

Furthermore, the difference of integrals~(\ref{eq:diffint}) has a finite part.
In order to reproduce the amplitudes (defined in a definite scheme, e.g. $MS$)
of full QCD in the effective theory, these parts, the \emph{matching
corrections}, have to be included in the renormalization constants.  In
ordinary effective theories such as HQET, such finite matching corrections will
not affect the scaling behavior to one-loop order, so they have to be taken
into account explicitly only from two-loop order on.  By contrast, in the
present case the finite matching corrections
\begin{equation}
  \Delta I_{\rm finite} =
\frac{i}{(4\pi)^{2}}\frac{1}{2v\cdot Q}
\ln^{2}
\biggl(
\frac{k_+-k_+^\prime-i\epsilon}{\mu}
\biggr)
\end{equation}
explicitly depend on the renormalization scale~$\mu$ and therefore modify the
scaling behavior already at one-loop order\footnote{The appeareance of nonlocal
divergencies and double logarithms in the matching corrections spoils the
naive formulation of a large-energy effective theory (LEET) for
\emph{exclusive} processes, as has been observed in \cite{Agli,BMS}.}.
 
In order to find the renormalization kernel, we take the imaginary 
part of  the difference of vertex corrections to the correlators 
$T$ and $\hat T$ resulting
from (\ref{eq:diffint}), keeping only UV--poles and double logarithms:
\begin{eqnarray}
\Delta Q^{\mu\nu} &=& \frac{1}{\pi} \mathrm{Im} 
(V_{\rm (sing)}^{\mu\nu} - \hat V_{\rm (sing)}^{\mu\nu}) \\
&=& 
\frac{\alpha_{s}}{\pi}\,
\frac12 {\rm Tr}\{\Gamma^{\mu\dagger}\fmslash{Q}\Gamma^\nu P_v^+\}
\int dk_{+}\delta(Q^{2} + 2v\cdot Q \,k_{+})
\\
&&\times
\int dk_{+}^{\prime}
\biggl[-Z^{(1)}(k_{+},k_{+}^{\prime},\mu)_{MS^{\prime}} 
- \frac{C_{F}}{\epsilon}\delta(k_{+}-k_{+}^{\prime} )\biggr] 
f^{(bare)}(k_+')\nonumber
\end{eqnarray}
The second term in square brackets accounts for wave function
renormalization of the heavy--quark fields. The one-loop contribution to 
vertex renormalization is given by
\footnote{The renormalization prescription which matches QCD in the
$MS$ scheme is denoted as $MS^\prime$.}
\begin{eqnarray}
Z^{(1)}(k_{+},k_{+}^{\prime},\mu)_{MS^{\prime}}  &=&C_{F}\biggl(\frac{2}{\epsilon^{2}} 
+ \frac{c-1}{\epsilon} \biggr)
\delta(k_{+} - k_{+}^{\prime}) \nonumber\\
&&{}-C_{F}\frac{2}{\epsilon}
\frac{d}{d k_{+}^{\prime}}
[\Theta(k_{+}^{\prime} - k_{+})\ln(\frac{|k_{+}^{\prime} -
k_{+}|}{\mu})]\nonumber \\
\label{eq:renconstMS}
&&{}+ \frac{1}{2}C_{F}\frac{d}{d k_{+}^{\prime}}
[\Theta(k_{+}^{\prime} - k_{+})\ln^{2}(\frac{|k_{+}^{\prime} -
k_{+}|}{\mu})]
\,,
\end{eqnarray}
which has to be understood in the distribution sense.

In this way we can define a renormalized shape function 
\begin{equation}
f(k_{+},\mu) = \int
dk_{+}^{\prime}Z(k_{+},k_{+}^{\prime},\mu)_{MS^{\prime}} 
f^{(bare)}(k_{+}^{\prime })
\end{equation}
where the one loop renormalization kernel
reads
\begin{equation}
Z(k_{+},k_{+}^{\prime},\mu)_{MS^{\prime}} = \delta(k_{+}-k_{+}^{\prime})
+ (\frac{\alpha_{s}}{\pi})Z^{(1)}(k_{+},k_{+}^{\prime},\mu)_{MS^{\prime}}.
\end{equation}
In what follows it will be convenient to switch to the 
$\overline{MS}^{\prime}$--scheme setting 
\begin{equation}
\frac{1}{\epsilon} = \frac{1}{\hat{\epsilon}} - \frac{1}{2}\ln(4\pi)
                      + \frac{1}{2}\gamma_{E}
\end{equation}
and keeping terms proportional to $1/\hat{\epsilon}^{2}$ and 
$1/\hat{\epsilon}$ in the renormalization--kernel:
\begin{eqnarray}
Z^{(1)}(k_{+},k_{+}^{\prime},\mu)_{\overline{MS}^{\prime}}  &=&C_{F}\biggl(\frac{2}{\hat{\epsilon}^{2}} 
- \frac{c+1}{\hat{\epsilon}} \biggr)
\delta(k_{+} - k_{+}^{\prime}) \nonumber \\
&-&C_{F}\frac{2}{\hat{\epsilon}}
\frac{d}{d k_{+}^{\prime}}
[\Theta(k_{+}^{\prime} - k_{+})\ln(\frac{|k_{+}^{\prime} -
k_{+}|}{\mu})] \nonumber \\
\label{eq:renconstMSb}
&+& \frac{1}{2}C_{F}\frac{d}{d k_{+}^{\prime}}
[\Theta(k_{+}^{\prime} - k_{+})\ln^{2}(\frac{|k_{+}^{\prime} -
k_{+}|}{\mu})]
\end{eqnarray}

Then to one loop order
\begin{equation}
\label{eq:renhadten}
\hat Q_{\mu\nu} = \frac{1}{2} 
\mathrm{Tr}\{\Gamma_{\mu}^{\dagger}\fmslash{Q} \Gamma_{\nu} P_v^+ \}
\int dk_+ \delta(Q^2 + 2v \cdot Q k_+) f(k_+,M)
\end{equation}
is UV--finite. Here $M$ is a scale of the order $v\cdot Q$ which is 
a perturbative scale.   As a consequence of the finite renormalization
in (\ref{eq:renconstMSb}) also the correct collinear
singularity is reproduced.

\section{Evolution equation}
\label{sec:evolution}

As a consequence of the renormalization procedure
the structure function now depends on both the renormalization scale 
$\mu$ and the renormalization--scheme ($\overline{MS}^{\prime}$)  
and obeys the evolution equation
\begin{equation}
\label{eq:eveq}
\frac{d}{d\ln\mu}f(k_+,\mu) = \int dk_{+}^{\prime}
\Gamma(k_{+},k_{+}^{\prime},\mu) f(k_+^{\prime},\mu).
\end{equation}
The evolution kernel $\Gamma(k_{+},k_{+}^{\prime},\mu)$ in the
$\overline{MS}^{\prime}$--scheme\footnote{In what follows this scheme
dependence of the evolution kernel is understood.} is defined implicitly by
\begin{equation}
\int dk_{+}^{\prime\prime}\Gamma(k_{+},k_{+}^{\prime\prime},\mu)
Z(k_{+}^{\prime\prime} ,k_{+}^{\prime},\mu)_{\overline{MS}^{\prime }} 
= \frac{d}{d\ln\mu}Z(k_{+} ,k_{+}^{\prime},\mu)_{\overline{MS}^{\prime }} ,
\end{equation}
where
\begin{equation}
  \frac{d}{d\ln\mu} \equiv 
  \frac{\partial}{\partial\ln\mu}
  - 2\alpha_s\left(\frac{\alpha_s}{\pi}\beta_0 + \ldots\right)
  \frac{\partial}{\partial\alpha_s}
\end{equation}
with $\beta_0=(33-2n_f)/12$.  Thus, to one-loop order the evolution kernel is
given by the negative coefficient of the $1/\epsilon$--pole term in
$Z^{(1)}(k_{+} ,k_{+}^{\prime},\mu)_{\overline{MS}^{\prime }}$ and the partial
$\ln(\mu)$--derivative of its finite part\footnote{Note that the double pole in
$Z^{(1)}(k_{+} ,k_{+}^{\prime},\mu)_{\overline{MS}^{\prime }} $ cancels with
the explicit $\mu$--derivative of the single pole--term:
\begin{equation}
 \frac{d}{d\ln\mu}\frac{d}{d k_{+}^{\prime}}
[\Theta(k_{+}^{\prime} - k_{+})\ln(\frac{|k_{+}^{\prime} - k_{+}|}{\mu})] 
 =-\delta(k_{+}^{\prime} - k_{+})
\end{equation}}:
\begin{eqnarray}
\Gamma(k_{+},k_{+}^{\prime},\mu)
&=&
(\frac{\alpha_{s}(\mu)}{\pi})\Gamma^{(1)}(k_{+},k_{+}^{\prime},\mu) \\ 
&=& (\frac{\alpha_{s}}{\pi})C_F\biggl( 
\delta(k_+^{\prime} - k_+) + \frac{d}{d k_{+}^{\prime}}
[\Theta(k_{+}^{\prime} - k_{+})\ln(\frac{|k_{+}^{\prime} -
k_{+}|}{\mu})] \biggr)\nonumber
\end{eqnarray}
In order to solve the evolution equation we 
Fourier transform
both the structure function and the evolution kernel: 
\begin{eqnarray}
\tilde f(\xi,\mu) &=& \int \frac{dk_{+}}{2\pi} f(k_{+},\mu) e^{ik_{+}\xi}
\nonumber \\
\label{eq:wloop}
\tilde \Gamma^{(1)}(\xi,\mu) &=& \int
\frac{dk_{+}}{2\pi}\Gamma^{(1)}(k_{+},0,\mu) e^{ik_{+}\xi}\\
&=& \frac{C_F}{2\pi}\biggl\{1 -  \ln(|\xi \mu|) - i\frac{\pi}{2} 
(\Theta(\xi) -\Theta(-\xi) )
\biggr\}
\end{eqnarray}
Then the evolution equation (\ref{eq:eveq}) reads
\begin{equation}
\frac{d}{d\ln \mu} \tilde f(\xi,\mu)  = (\frac{\alpha_{s}}{\pi})
 (2\pi) \tilde \Gamma^{(1)}(\xi,\mu) \tilde f(\xi,\mu)
\end{equation} 
and can be solved easily
\begin{equation}
\tilde f(\xi,\mu) = \tilde U(\xi,\mu,\mu_{0}) \tilde f(\xi,\mu_{0}) 
\end{equation}
where
\begin{equation}
\tilde U(\xi,\mu,\mu_{0}) = K(\mu,\mu_{0}) 
\exp\left(i\frac{\pi}{2}\omega \varepsilon(\xi)\right)
 |\mu_{0} \xi|^{\omega}
\end{equation}
with
\begin{equation}
\omega = \frac{C_{F}}{2\beta_{0}} 
          \ln\biggl(\frac{\alpha_{s}(\mu)}{\alpha_{s}(\mu_{0})}\biggr),
\qquad
\varepsilon(\xi) = \Theta(\xi)-\Theta(-\xi)
\end{equation}
and
\begin{equation}
K(\mu,\mu_{0}) = \biggl(\frac{\mu}{\mu_{0}}\biggr)^{-\frac{C_F}{2\beta_{0}}}
 \biggl (\frac{\alpha_{s}(\mu)}{\alpha_{s}(\mu_{0})}\biggl)
^{\displaystyle{-\frac{C_F}{2\beta_{0}}[1 +
  \frac{\pi}{2\beta_{0}\alpha_{s}(\mu_{0})}]}}\,.
\end{equation}
Transforming back to momentum space we get
\begin{equation}
\label{eq:soleveq}
f(k_{+},\mu) = \int dk_{+}^{\prime} U(k_{+},k_{+}^{\prime};\mu,\mu_{0})
             f(k_{+}^{\prime},\mu_0)
\end{equation}
where
\begin{equation}
\label{eq:RGevsimp}
U(k_{+},k_{+}^{\prime};\mu,\mu_{0})
 =-K(\mu,\mu_{0}) \Gamma(1 + \omega) 
 \mu_{0}^{\omega}
\frac{\sin \pi \omega}{\pi} 
\frac{\Theta(k_+^{\prime} - k_+ )} 
{(k_+^{\prime} - k_+)^{1 + \omega}}\,.
\end{equation}
Note that the tree-level result is recovered
from  (\ref{eq:RGevsimp}) by letting $\mu =\mu_0$
and performing the limit $\omega \rightarrow -0$
carefully, since 
\begin{equation}
\lim_{\omega \rightarrow -0}\frac{\sin \pi \omega}{\pi}
\frac{\Theta(k_+^{\prime}
-k_+)}{(k_+^{\prime}
-k_+)^{1+\omega}} =-\delta(k_+^{\prime} -k_+)\, .
\end{equation} 
 
Once radiative corrections in the 
definition of the light--cone structure function are included, we 
are in the position to study the invariant mass
spectrum.
To this end we insert (\ref{eq:soleveq}) into (\ref{eq:renhadten}),
choose for $\mu$ a high energy scale $M=\mathcal O(v\cdot Q)$ and 
let $\mu_0 = \mu$
\begin{equation}
\hat Q_{\mu\nu} = \frac{1}{2} 
\mathrm{Tr}\{\Gamma_{\mu}^{\dagger}\fmslash{Q} \Gamma_{\nu} P_v^+ \}
 \int dk_+^{\prime}
U(k_+,k_+^{\prime};M,\mu)f(k_+^{\prime},\mu)
\end{equation}
The coefficient function $U(k_+,k_+^{\prime};M,\mu)$ is analogous to the set of
Wilson coefficients $C_i(M,\mu)$ in local effective field theories.

Note that, by definition, the convolution of the renormalized structure
function with the coefficient function is independent of the 
choice of the renormalization scale $\mu$.
At the low energy scale $\mu = \mathcal O(\Lambda_{QCD})$ the structure 
function is only weakly affected by perturbative corrections and 
can be safely considered as a pure non--perturbative object
which may be described by a model~\cite{models}:
\begin{equation}
f(k_+,\Lambda_{QCD}) = f^{(model)}(k_+)
\end{equation}


Putting everything together, the renormalization group
improved hadronic tensor becomes:
\begin{eqnarray}
\hat Q_{\mu\nu} &=&  -\frac{1}{2(2 v\cdot Q)} 
 K(M,\mu) \Gamma(1 + \omega) 
 \mu^{\omega}
\frac{\sin \pi \omega}{\pi} 
\mathrm{Tr}\{\Gamma_{\mu}^{\dagger}\fmslash{Q} \Gamma_{\nu} P_v^+ \}\\
& &\int dk_+^{\prime} 
\frac{\Theta(k_+^{\prime} - k_+)}{(k_+^{\prime} - k_+)^{1 + \omega}}
f^{(model)}(k_+^{\prime})
\end{eqnarray}


\section{Evolution of Moments}
It is well known that the moments of the light--cone structure function 
may be related to matrix elements of local operators. The moments 
satisfy the relation
\begin{equation} \label{eq:mom}
M_n = \int dk_+ (k_+)^n f(k_+) = 
      \langle B(v) | \bar{h}_v (iD_+)^n h_v | B(v) \rangle 
\end{equation}
in particular we have $M_0 = 1$ and $M_1 = 0$ due to the normalization 
and the equations of motion. 

However, since the light--cone distribution funtion evolves with a 
change of scale, also the moments depend on the scale. The moments 
at the scale $\mu$ are given by 
\begin{equation} 
M_n (\mu) = \int dk_+ (k_+)^n f(k_+,\mu) = 
            \int dk_+ (k_+)^n \int dk_+^\prime 
            U(k_+, k_+^\prime, \mu, \mu_0) f(k_+^\prime ,\mu_0)
\end{equation}
Since the kernel $U (k_+, k_+^\prime, \mu, \mu_0)$ depends only 
on the difference of $k_+$ and $k_+^\prime$ one may reexpress 
these moments in terms of the moments at scale $\mu_0$ by a simple 
change of variables. One obtains
\begin{equation} \label{momrel}
M_n (\mu) = \sum_{j=0}^n \frac{n!}{j! \, (n-j)!} M_j (\mu_0) 
            \int dk_+ (k_+)^{n-j} U(k_+, 0, \mu, \mu_0)   
\end{equation}
which means that under renormalization the $n^{th}$ moment will 
depend on all the moments $M_j$ with $j \le n$ at the lower scale. 

However, due to the power type behaviour in the variable 
$k_+ - k_+^\prime$ of the evolution kernel 
$U(k_+, k_+^\prime, \mu, \mu_0)$ the integral in (\ref{momrel}) 
does not converge, since the region of integration is 
$-\infty \le k_+ \le 0$ and hence it has to be regularized. 
If we regularize the integral by restricting the integration 
region to $-\Lambda \le k_+ \le 0$ with some cut off $\Lambda$,
the dependence on $\Lambda$ can be already guessed from dimensional 
analysis. More precisely, the moments $M_n (\mu)$ exhibit power 
divergencies where the dependence on $\Lambda$ is given by
\begin{equation} 
M_n (\mu) = \sum_{j=0}^n M_j (\mu_0) 
            C_{n,j} (\mu, \mu_0) 
            \left( \frac{\mu_0}{\Lambda} \right)^\omega 
            \Lambda^{n-j} 
\end{equation}
where from (\ref{eq:RGevsimp}) we get  
\begin{equation}
C_{n,j} (\mu, \mu_0) = 
-\frac{n!}{j! \, (n-j)!}K(\mu,\mu_{0}) \Gamma(1 + \omega) 
\frac{\sin \pi \omega}{\pi}\,.
\end{equation}
To interpret this result it is instructive to consider the first few 
moments. The normalization receives a multiplicative renormalization 
\begin{equation}
M_0 (\mu) = M_0 (\mu_0) 
            C_{0,0} (\mu, \mu_0) 
            \left( \frac{\mu_0}{\Lambda} \right)^\omega 
\end{equation}
while the first moment receives contributions from both $M_0 (\mu_0)$ and
$M_1 (\mu_0)$:
\begin{equation}
M_1 (\mu) = M_1 (\mu_0) 
            C_{1,1} (\mu, \mu_0) 
            \left( \frac{\mu_0}{\Lambda} \right)^\omega 
            + 
            \Lambda M_0 (\mu_0) C_{1,0} (\mu, \mu_0)
            \left( \frac{\mu_0}{\Lambda} \right)^\omega 
\end{equation}
However, the first moment is 
\begin{equation}
M_1 =  \langle B(v) | \bar{h}_v (iv\cdot D) h_v | B(v) \rangle 
\end{equation}
which according to the equations of motion should vanish. This is true, 
if the pole mass of the heavy quark is taken as the expansion parameter, 
such that no residual mass term appears. The choice of a different mass 
definition $m \to m + \delta m$ will modify the equation of motion into
\begin{equation}
(iv\cdot D) h_v = \delta m h_v 
\end{equation}
and hence the running of the first moment implies a change in the 
pole mass definition due to the power divergence 
\begin{equation} 
\delta m = \Lambda C_{1} (\mu, \mu_0) 
           \left( \frac{\mu_0}{\Lambda} \right)^\omega\,.
\end{equation}
In principle this relation can be used to fix the value of the cut-off 
in terms of the pole mass; once this is done, all higher moments can 
be computed in terms of the pole mass at the lower scale $\mu_0$.  

Although the shape function is an object entirely defined in HQET and 
hence should be independent of the mass $m_b$, in this indirect way a 
mass dependence comes into the game, at least if we insist to interpret 
the moments in terms of the matrix elements (\ref{eq:mom}).  

\section{Conclusions}
\label{sec:conclusions}
In this paper we have proposed a method to combine 
perturbative and non-perturbative contributions
to hadronic invariant mass spectra in the endpoint region
where the invariant mass $Q_H^2$ becomes small of the order
$\mathcal O(\Lambda_{QCD}m_B)$.
At tree level the non--perturbative corrections of leading twist are
resummed into  a universal light--cone distribution function.
However, taking radiative corrections into account,
this function has to be renormalized. 
In order to match the 
leading IR--singularity of the QCD spectrum,
we have chosen a renormalization scheme
which except standard UV--renormalization
provides corrections of the IR--behaviour of the 
structure function adding proper finite terms to the 
renormalization kernel.
As a consequence of renormalization the renormalization scale dependence of 
the structure function is described by an evolution equation.
We have computed the evolution kernel to one loop order,
which deviates from the usual $\overline{MS}$--kernel, since  
we have included a finite renormalization in order to reproduce the 
infrared behaviour of full QCD.

The analytical solution of the evolution equation 
yields a resummation of logarithmic corrections and
relates structure functions at 
different renormalization scales.  
The last property has then be used to include 
radiative corrections in the invariant mass spectrum
in a manner common from usual renormalization group techniques.

As the structure function itself is scale dependent so are its
moments. However, including radiative corrections the 
moments exhibit an UV--di\-ver\-gence which has been regularized by an hard cut off
thereby mixing moments of different order.
We fixed the cut off by relating the first moment to 
the definition of the pole mass.
Once the value of the cut off is known, the scale dependence of 
all other moments is computable.

\section*{Acknowledgements}

We are grateful to M.~Neubert and 
V.~Smirnov for valuable discussions.
This work was supported by 
the ``Graduiertenkolleg: Elementarteilchenphysik and Beschleunigern'' 
and the ``Forschergruppe: Quantenfeldtheorie, Computeralgebra und 
Monte Carlo Simulationen'' of the 
Deutsche Forschungsgemeinschaft.   
W.K.\ is supported by German Bundesministerium f\"ur Bildung und
Forschung (BMBF), Contract No.~05~6HD~91~P(0).


\begin{thebibliography}{10}

\bibitem{FaLuSav1}
 A.F. Falk, M. Luke and M.J. Savage, Phys. Rev. {\bf D53}, 2491 (1996).
\bibitem{FaLuSav2}
 A.F. Falk, M. Luke and M.J. Savage, Phys. Rev. {\bf D53}, 6316 (1996).
\bibitem{FaLu}
 A.F. Falk and M. Luke, Phys. Rev. {\bf D57}, 424 (1998)
\bibitem{Neu}
 M. Neubert, Phys. Rev. {\bf D49}, 3392 and 4623 (1994).
\bibitem{ManNeu}
 T. Mannel and M. Neubert, Phys. Rev. {\bf D50}, 2037 (1994).
\bibitem{Bigietal}
 I.I. Bigi, M.A. Shifman, N.G. Uraltsev and A.I. Vainshtein,
Phys. Lett. {\bf B328}, 431 (1994)
\bibitem{KorSter}
 G.P. Korchemsky and G. Sterman, Phys. Lett. {\bf B340}, 96 (1994).
\bibitem{NeuGroz}
  A.G. Grozin and M. Neubert, Phys. Rev. {\bf D55}, 272 (1997).
\bibitem{KorGroz}
  A.G. Grozin and G.P. Korchemsky, Phys. Rev. {\bf D53}, 1378 (1996).
\bibitem{LEET}
 M.J. Dugan and B. Grinstein, Phys.Lett. {\bf B255}, 583 (1991).
\bibitem{Agli}
 U. Aglietti and G. Corbo, hep-ph 9803485.
\bibitem{BMS}
 C. Balzereit, T. Mannel and V. Smirnov, in preparation.
\bibitem{models}
 \emph{The BABAR Physics Book},
 SLAC--R--504, in preparation.
\end{thebibliography}
\end{document}